\documentstyle[11pt,newpasp,twoside,epsf]{article}
\def\edcomment#1{\iffalse\marginpar{\raggedright\sl#1\/}\else\relax\fi}
\reversemarginpar

\begin{document}
\title{Extragalactic water masers in bright IRAS sources}
 \author{A.\ Tarchi$^{1,2}$, C.\ Henkel$^{3}$, A.B.\ Peck$^{4}$, N.\ Nagar$^{5}$, L.\ Moscadelli$^{2}$, and K.M.\ Menten$^{3}$}
\affil{$^{1}$Instituto di Radioastronomia, CNR, Bologna, Italy}
\affil{$^{2}$Osservatorio Astronomico di Cagliari, Capoterra (CA), Italy}
\affil{$^{3}$MPIfR, Bonn, Germany}
\affil{$^{4}$Harvard-Smithsonian CfA, SMA Project, Hilo, HI, USA} 
\affil{$^{5}$Kapteyn Astronomical Institute, Groningen, the Netherlands}

\begin{abstract} 
We report the results of a search for 22\,GHz water maser emission in IRAS-bright galaxies, using the 100-m Effelsberg telescope. In particular, we present the details of four new maser detections (IC~342, NGC~2146, NGC~3556, and Arp~299) and follow-up interferometric studies. A comparison between water maser detection rates derived in the present study and those in previous similar surveys is also presented.
\end{abstract}

\section{Introduction}

To date, there is evidence for a total of three distinct classes of extragalactic H$_2$O masers: i) H$_2$O megamasers (with isotropic luminosities L$_{\rm iso}$ $>$\,10\,L$_{\odot}$) associated with accretion disks in active galaxies (e.g.\ NGC~4258, Miyoshi et al.\ 1995) ii) H$_2$O megamasers resulting from an interaction between the nuclear radio jet and an encroaching molecular cloud (e.g. Mrk~348, Peck et al.\ 2001, 2003) iii) weaker H$_2$O masers, the `kilomasers' (with L$_{\rm iso}$ $<$\,10\,L$_{\odot}$), often associated with prominent star forming regions in large scale galactic disks, and, thus far found in galaxies containing bright IRAS (Infrared Astronomical Satellite) point sources (e.g.\ NGC~2146, Tarchi et al.\ 2002b).

We have undertaken deep searches optimized to detect emission arising from the last class of sources. The sample (hereafter `FIR-maser' sample) is comprised of all galaxies with declination $>$--30{\deg}, known velocity, and IRAS point source flux densities of $S_{100\mu \rm m}$$>$50\,Jy (e.g. IRAS 1989). There is a total of 45 sources.

\section{Observations}
The target sources of the sample were measured in the $6_{16} - 5_{23}$ line of H$_2$O (rest frequency: 22.23508\,GHz) with the 100-m telescope of the MPIfR at Effelsberg on various occasions between April 2001 and September 2002. The full width at half power beamwidth was $\sim$40\arcsec\ and the pointing accuracy was always better than 10\arcsec. The autocorrelator backend was split into eight bands of width 40 or 80\,MHz and 512 or 256 channels each. This yielded channel spacings of $\sim$1 or $\sim$4\,km\,s$^{-1}$.

\section{Results}

\begin{figure}
\plotone{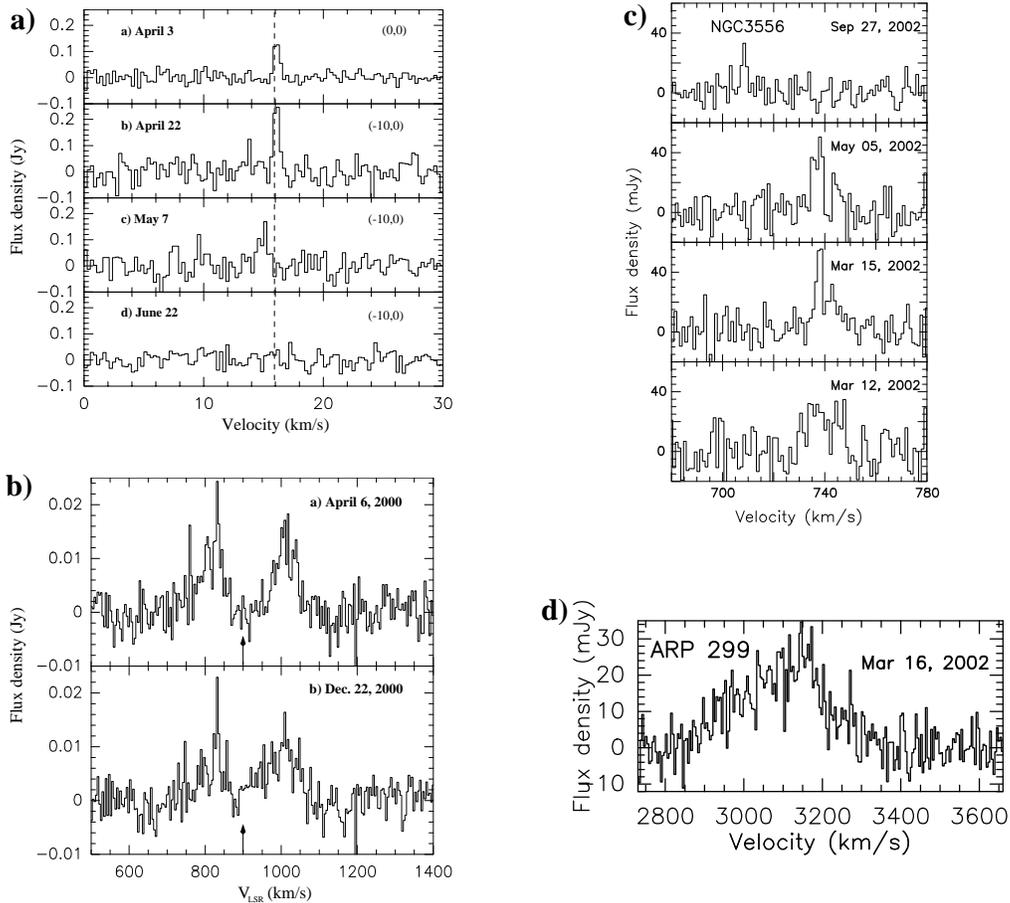}
\caption{H$_2$O kilomaser spectra observed with the 100-m telescope of the MPIfR at Effelsberg toward: {\bf {a)}} IC~342 {\bf {b)}} NGC~2146; {\bf {c)}} NCG~3556; {\bf {d)}} Arp~299.}
\end{figure}

Including the early part of our survey (Tarchi et al.\ 2002a, b), from the FIR--sample, we have detected three new kilomasers (IC~342, NGC~2146, and NGC3556) and a new megamaser (Arp~299). Line profiles are shown in Fig.\,1. Properties of the individual sources are discussed below.

\subsection{Recent detections}

{\bf {IC~342:}} We have obtained the first definite detection of water maser emission in the nearby spiral galaxy IC~342 (at a distance $D$ = 1.8 Mpc). The maser arises from a location 10--15$''$ to the west of the centre of the galaxy\footnote{The position has been derived from an Effelsberg map we have produced taking advantage of the high line intensity and good weather conditions} and is associated with a powerful star forming region at a projected distance of $\sim$ 100 pc from the nucleus. Time variability (see Fig.\,1{\bf {a}}), if intrinsic, yields a maser size of $\rm \la 1.5 \times 10^{16}\:cm^{-3}$ and a brightness temperature $\rm \ga 10^{9} \:K$. 

\noindent {\bf {NGC~2146:}} The one detected toward the starburst galaxy NGC~2146 ($D$ = 14.5 Mpc) is the most luminous and distant H$_{2}$O kilomaser detected so far (see Fig.\,1{\bf {b}}). Interferometric observations with the Very Large Array (VLA) show that a part of the emission originates from two prominent sites of star formation that are associated with compact radio continuum sources, likely ultra-compact {H\kern0.1em{\sc ii}} regions.

\noindent {\bf {NGC~3556:}} is a nearby spiral galaxy located at a distance of $\sim$12\,Mpc. Its FIR luminosity, $L_{\rm FIR}$ $\sim$ 10$^{10}$\,L$_{\odot}$, is similar to that of the Milky Way. The detected H$_{2}$O maser line initially had a central velocity of $\sim$738\,km\,s$^{-1}$. With a peak flux of 20--40\,mJy, the maser had an isotropic luminosity of $\sim$1\,L$_{\odot}$. More recently, the maser feature disappeared and another weaker component, at $\sim$708\,km\,s$^{-1}$, was detected in September 2002. The profiles are shown in Fig.\,1{\bf {c}}.

\noindent {\bf {Arp~299:}} is a merging system composed of three main regions of activity: the nuclear regions of IC~694 in the east, NGC~3690 in the west, and the interface where IC~694 and NGC~3690 overlap, located approximately 10\arcsec\ north of NGC~3690. In Arp~299, the water maser profile is extremely broad ($\sim$200\,km\,s$^{-1}$), with a peak flux density of 30\,mJy (Fig.\,1{\bf {d}}). Adopting a distance of 42\,Mpc, the total isotropic luminosity is $\sim$250\,L$_{\odot}$, placing the object among the more luminous H$_2$O megamaser sources. The maser line is centered at a velocity of 3100\,km\,s$^{-1}$, i.e. very close to the systemic velocity of the entire complex of sources which constitute Arp~299.

\section{Discussion and Conclusions}

More than 700 active galaxies have been observed in water vapor maser surveys to date. In order to search for a large number of sources, these surveys were not very sensitive and had a low detection rate, between zero (e.g. Henkel et al.\ 1998) and a few percent (e.g.\ Henkel et al.\ 1984; Braatz, Wilson, \& Henkel 1996, Greenhill et al.\ 2002).

We detected with IC~342, NGC~2146, NGC~3556 and Arp~299 four new H$_2$O masers. The new detections are a consequence of higher sensitivity (noise levels of $\sim$ 10\,mJy for a 1\,km\,s$^{-1}$ channel), more stable continuum sampling over large bandwidths (i.e. improved baselines), and luck (in the case of the short-lived flare observed toward IC~342). Including all previously detected sources in the complete FIR--sample, we find a detection rate of 10/45 (or 22\%).

\begin{figure}
\plotfiddle{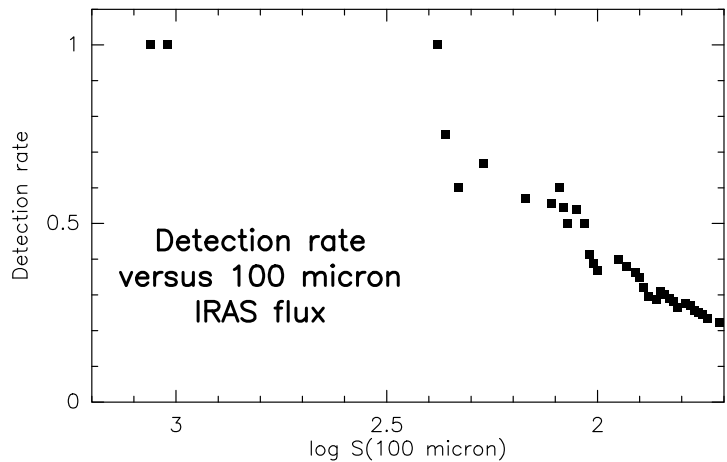}{6cm}{0}{80}{80}{-130}{0}
\caption{Detection rate of the H$_2$O FIR-maser sample (see Sect.\,1 for selection criteria) for all galaxies above a given IRAS Point Source Catalog $S_{\rm 100\mu m}$ flux density.}
\end{figure}

The high rate of maser detections in our sample of galaxies strongly suggests that a relationship between FIR flux density and maser phenomena exists, consistent with the assessment by Henkel, Wouterloot, \& Bally (1986). Fig.\,2 shows the cumulative detection rate above a given 100$\mu$m IRAS Point Source Catalog flux for the parent galaxy. The detection rate strongly declines with decreasing FIR flux. For fluxes $\sim$1000\,Jy, 100--300\,Jy, and 50--100\,Jy, we find detection rates of 2/2 (or 100\%), 5/17 (or 29\%), and 3/26 (or 12\%). This indicates that an extension of the sample to lower fluxes would be worthwhile. 

It appears that $S_{\rm 100\mu m}$ (and thus also the FIR luminosity, $L_{\rm FIR}$) and H$_2$O peak fluxes are roughly proportional, as was already suggested by Henkel et al.\ (1986) on the basis of a smaller number of detected sources. 

The possibility that a correlation between $L_{\rm IR}$ and $L_{\rm H_2O}$, similar to that found by Genzel \& Downes (1979) for galactic star forming regions, exists is under investigation.

\end{document}